\pdfoutput=1

\documentclass[pdf,a]{iucr}              

\usepackage{amssymb}
\usepackage{graphicx}
\usepackage{amsmath}

     \paperprodcode{a000000}      
     \paperref{xx9999}            
     \papertype{FA}               

     \paperlang{english}          
     \journalcode{J}              
     \journalyr{2009}
     \journaliss{0}
     \journalvol{0}
     \journalfirstpage{000}
     \journallastpage{000}
     \journalreceived{0 XXXXXXX 0000}
     \journalaccepted{0 XXXXXXX 0000}
     \journalonline{0 XXXXXXX 0000}

\begin{document}                  



\title{Experimental and theoretical study of diffraction properties of various crystals for the realization of a soft gamma-ray Laue lens}

\shorttitle{Experimental and theoretical study of crystals for a Laue lens}


\cauthor[a]{Nicolas}{Barri{\`e}re}{nicolas.barriere[at]iasf-roma.inaf.it}{address if different from \aff}
\author[b]{Julien}{Rousselle}
\author[b]{Peter}{von Ballmoos}
\author[c]{Nikolai V.}{Abrosimov}
\author[d]{Pierre}{Courtois}
\author[e]{Pierre}{Bastie}
\author[b]{Thierry}{Camus}
\author[d]{Michael}{Jentschel}
\author[f]{Vladimir N.}{Kurlov}
\author[a]{Lorenzo}{Natalucci}
\author[b]{Gilles}{Roudil}
\author[g]{Nicolai}{Frisch Brejnholt}
\author[h]{Denis}{Serre}

\aff[a]{INAF - IASF Roma, via Fosso del Cavaliere 100, 00130 Roma \country{Italy}}
\aff[b]{CESR - UMR 5187, 9 Av. du Colonel Roche, 31028 Toulouse \country{France}}
\aff[c]{IKZ, Max Born-Str. 2, D-12489 Berlin \country{Germany}}
\aff[d]{ILL, 6 rue Jules Horowitz,  38042 Grenoble \country{France}}
\aff[e]{LSP Ð UMR 5588, 140 Av. de la physique, 38402 Saint Martin d'H{\`e}res \country{France}}
\aff[f]{Institute of Solid State Physics of Russian Academy of Sciences, 142432 Chernogolovka \country{Russia}}
\aff[g]{DTU Space, Juliane Maries Vej 30, 2100 Copenhagen \country{Denmark}}
\aff[h]{Leiden Observatory, Leiden University, P.O. Box 9513, 2300 RA Leiden \country{The Netherlands}}


\shortauthor{Barri{\`e}re, N. et al.}







\maketitle                        

\begin{synopsis}
Supply a synopsis of the paper for inclusion in the Table of Contents.
\end{synopsis}

\begin{abstract}
Crystals are the elementary constituents of Laue lenses, an emerging technology which could allow the realization of a space borne telescope 10 to 100 times more sensitive than existing ones in the 100 keV - 1.5 MeV energy range. This study addresses the current endeavor to the development of efficient crystals for the realization of a Laue lens.

In the theoretical part 35 candidate-crystals both pure and two-components are considered. Their peak reflectivity at 100 keV, 500 keV and 1 MeV is calculated assuming they are mosaic crystals. It results that a careful selection of crystals can allow a reflectivity above 30\% over the whole energy range, and even reaching 40\% in its lower part.

Experimentally, we concentrated on three different materials (Si$_{1-x}$Ge$_x$ with gradient of composition, mosaic Cu and Au) that have been measured both at ESRF and ILL using highly-monochromatic beams ranging from 300 keV up to 816 keV. The aim was to check their homogeneity,  quality and angular spread (mosaicity). These crystals have shown outstanding performance such as reflectivity up to 31\% at  $\sim$ 600 keV (Au) or 60\% at 300 keV (SiGe) and angular spread as low as 15 arcsec for Cu, fulfilling very well the requirements for a Laue lens application. Unexpectedly, we also noticed important discrepancies with Darwin's model when a crystal is measured using various energies. \\
\end{abstract}


\section{Introduction}

Despite the very rich physics it offers, the soft gamma-ray sky is not fully exploited at present because the telescopes in this domain are blinded by intense and complex instrumental background induced in their detectors by space environment (cosmic rays, radiations belts, Earth albedo and solar ßares, see e.g. [Weidenspointner et al., 2005]). Current instruments operating in this part of the electromagnetic spectrum do not use focusing optics. They reconstruct the incidence direction of detected events thanks to either an aperture modulation (coded mask) or by tracking the multiple (Compton) interactions of photons in a sensitive volume. The common point of these two techniques is that the signal is collected onto an area which is itself the sensitive area. 

To keep improving our knowledge of the violent celestial processes responsible of the emission of high-energy photons we need to develop more sensitive telescopes. With the existing kind of telescopes more sensitive means larger in order to collect more signal. But the improvement of sensitivity only scales with the square root of the collection surface since the instrumental background scales with the volume of detectors. This is why it appears impossible to make the required sensitivity leap of a factor 10-100 with the existing principles of soft gamma-ray telescopes.

A new approach involving the concentration of gamma rays has been studied for more than a decade and proved to be feasible in the $\sim$100  keV - 1.5 MeV domain [von Ballmoos et al., 2004]. This paper addresses the endeavour that has been ongoing for the last four years to develop efficient elementary constituents for such lens: crystals.

The principle of Laue lenses is described in the next section. The third section gives an overview of theories used to model diffraction in both mosaic crystals and crystals having curved diffracting planes. The fourth section deals with the theoretical study that has been undertaken to identify among pure-element and two-component crystals the best suited  for a use in a Laue lens. The fifth section is devoted to the presentation of experimental characterizations of the three different types of crystal of which diffraction properties have been investigated: gradient silicon-germanium (SiGe), copper (Cu) and gold (Au). Analysis of experimental results and comparison with theoretical predictions are given in subsections associated with each crystal type. The last subsection emphasizes the discrepancies that we have noticed with mosaic crystals between Darwin's model and experimental results. The conclusions and perspectives of this work are given in the last section.

\section{Principle of a Laue lens}

\begin{figure}
\includegraphics[width=0.8\textwidth]{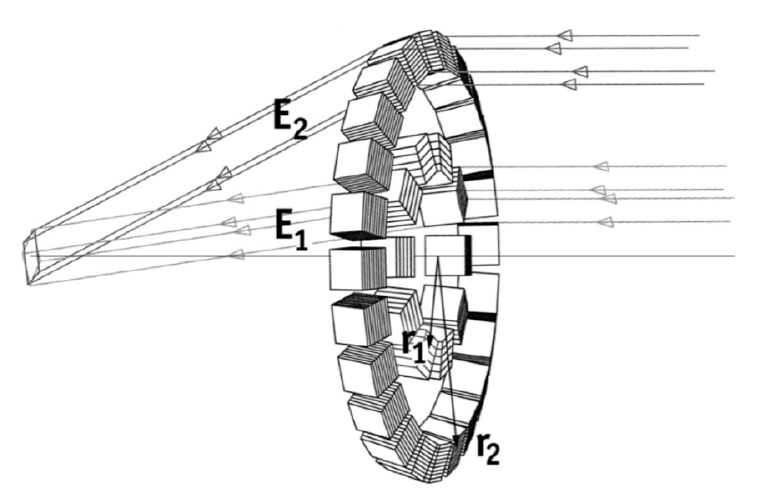}
\caption{Principle of a Laue lens: A large number of crystal tiles arranged in concentric rings diffract radiations coming from infinity towards a common focus. Depending on the radius of rings and the d-spacing of the crystals, several rings can concentrate the same energy or not. }
\label{fig:LauelensPrinciple}
\end{figure}

A Laue lens concentrates gamma-rays using Bragg diffraction in the volume of a large number of crystals arranged in concentric rings and accurately orientated in order to diffract radiation coming from infinity towards a common focal point (e.g. [Lund, 1992]). In the simplest design each ring is composed of identical crystals, their axis of symmetry defining the line of sight of the lens (see Figure \ref{fig:LauelensPrinciple}).

Bragg's law links the angle $\theta_B$ between the rays direction of incidence and the diffraction planes to the diffracted energy $E$ through the diffracting planes d-spacing $d_{hkl}$:

\begin{eqnarray}
2d_{hkl} \sin \theta_B = {hc \over E} \\
\Leftrightarrow \: 2d_{HKL} \sin \theta_B = n\, {hc \over E} \notag
\end{eqnarray}
with $h$, $k$, $l$ the Miller index defining the set of diffracting planes at work (another notation uses $H$, $K$, $L$ prime numbers and $n$ the order of diffraction), $h$ the Planck constant and $c$ the velocity of light in vacuum. Considering a focal distance $f$, the mean energy diffracted by a ring only depends on its radius $r$ and the d-spacing of its constituent crystals [Halloin \& Bastie, 2005]:
\begin{eqnarray}
E = {hc \over 2 d_{hkl} \sin \left({1 \over 2} \arctan \left({r \over f} \right) \right)} \varpropto {f \over d_{hkl} \, r}
\end{eqnarray}

We can distinguish two types of lenses. The first one uses either various crystalline materials or various reflections (keeping in mind that higher orders are less efficient) to maintain the product $d_{hkl}\,r$ constant in every ring so that they all diffract the same energy ($E_1 = E_2$ in Figure \ref{fig:LauelensPrinciple}). This principle allows the realization of a narrow bandpass lens like the prototype CLAIRE which used eight Ge reflections in eight rings to focus around 170 keV [Halloin et al., 2003; von Ballmos et al., 2004].

On the other hand if the same reflection of a given material is used on many consecutive rings, the diffracted energy is slightly shifted from one ring to the next increasing towards smaller radii ($E_1 > E_2$ in Figure \ref{fig:LauelensPrinciple}). If the crystals diffract a bandpass large enough, the contributions diffracted by neighboring rings overlap resulting in a broadband continuous energy coverage.

Recent projects of Laue lens telescope combine both effects. They use several ranges of rings (each range being composed of rings using a single reflection) that superimpose to cover efficiently broad energy bands. One such project, the Gamma Ray Imager (GRI) that has been proposed\footnote{in June 2007, as an answer to the first announcement of opportunity of the long term plan Cosmic Vision 2015-2025} to the European Space Agency (ESA), uses mainly Cu 111, Cu 200, Cu 220 and SiGe 111 to create two broad bandpasses ranging over 220 - 650 keV and 790 - 910 keV achieving several hundreds of cm$^2$ of effective area [Knodlseder et al., 2007; Barri{\`e}re et al., 2007].

Single perfect crystals only diffract an energy band  of a few eV wide, given by the Darwin width (see e.g. [Authier, 2001]). A Laue lens requires that crystals diffract a larger band. Two types of crystals can achieve this: mosaic crystals and crystals having curved diffracting planes (CDP crystals). For a GRI-like lens crystals optimal angular spread (hereafter called mosaicity\footnote{for the sake of simplicity, we propose to call mosaicity (noted $\Omega$) the full width at half-maximum of angular distribution of diffracting planes, even in the case of crystals with curved planes \label{fn:mosaicity}}) has been proven to be around 30 arcsec [Barri{\`e}re, 2008], a figure which should be multiplied by a factor of $\sim$2 - 3 for Laue lenses having a shorter focal length (GRI has a focal length of 100m).

\section{Theory of diffraction in mosaic and CDP crystals}

\subsection{Definitions}

We shall introduce two quantities that will be used in the following. The \emph{reflectivity} is defined as the ratio of the diffracted beam intensity over the incident beam intensity. And the \emph{diffraction efficiency} is defined as the ratio of the diffracted beam intensity over the intensity of the transmitted beam when no diffraction occurs.

\subsection{Mosaic crystals}
\label{sec:MosCryst}

Mosaic crystals are described using Darwin's model as an assembly of tiny identical small perfect crystals, \textit{the crystallites}, each slightly misaligned with respect to the others according to an angular distribution usually taken as Gaussian [Darwin, 1914, 1922]. Zachariasen [1945] gives the equation of the intensity diffracted as a function of the angle of incidence:
\begin{eqnarray}
I_{h, mos} = I_0 \, {1\over 2}{\left( 1 - e^{ -2 \sigma T_0} \right) \, e^{-\mu T_0 \over \cos \theta_B}}
\label{eq:I_h_mos}
\end{eqnarray}
where $I_0$ is the incident intensity, $T_0$ is the crystal thickness, $\mu$ is the linear absorption coefficient, and $\sigma$ can be interpreted as the coherent diffusion coefficient that is written as:
\begin{eqnarray}
\sigma = W(\Delta \theta) Q
\label{eq:def_sigma}
\end{eqnarray}
where $\Delta \theta$ is the difference between the actual angle of incidence of the beam onto the diffracting planes and Bragg angle. $W$ is the distribution function of the crystallites orientation and $Q$ is the integrated intensity diffracted by a single perfect crystal per unit of thickness. $Q$ is given by the dynamical theory of diffraction:
\begin{eqnarray}
Q_{dyn} = {\pi^2 \, d_{hkl} \over \Lambda_0^2 \, \cos \theta_B} \, f(A)
\label{eq:defQ}
\end{eqnarray}
where in the Laue symmetric case $f(A)$ is given by:
\begin{eqnarray}
f(A) &=& {B_0(2A) + |\cos2\theta_B| \,  B_0(2A|\cos2\theta_B|) \over 2A(1+ \cos^2\theta_B)} \\
 & \approx& {B_0(2A) \over 2A}. \label{eq:f(A)_approx} \label{eq:fA_approx}
\end{eqnarray}
The above approximation (\ref{eq:fA_approx}) can be done for $\theta_B$ small, which is always valid for energies above 100 keV. $B_0$ is the Bessel function of zero order integrated between 0 and $2A$ and $A$ is defined as
 \begin{eqnarray}
 A = {\pi \, t_0 \over \Lambda_0 \, \cos \theta_B},
 \label{eq:defA}
 \end{eqnarray}
in which $t_0$ is the crystallites thickness. $\Lambda_0$ is called the \emph{extinction length} and is defined for the Laue symmetric case (see e.g. [Authier, 2001]) as:
\begin{eqnarray}
\Lambda_0 = {\pi\, V_c \, \cos \theta_B \over r_e \lambda \, |C| \, |F_{hkl}| },
\label{eq:extlength}
\end{eqnarray} 
with $F_{hkl}$ being the structure factor (taking into account the electrons repartition in space, the crystal lattice and the effect of temperature via the so-called Debye-Waller's factor which is included in the structure factor), $V_c$ the volume of the crystal cell, and $C$ the polarization factor. 

We note that the dynamical theory tends towards the kinematical theory when $t_0 \ll \Lambda_0$
(which implies $f(A) \rightarrow 1$ in (\ref{eq:defQ})). In this case crystals are referred to as \emph{ideally imperfect} crystals and $Q$ is given by $Q_{kin}={\pi^2 \, d_{hkl} \over \Lambda_0^2 \, \cos \theta_B}$.

The distribution function of the crystallites orientation $W$ can be expanded as:
\begin{eqnarray}
W(\Delta \theta) = 2 \sqrt{\ln(2) \over \pi}\, {1\over \Omega} \, e^{-ln(2)\left( {\Delta \theta \over \Omega / 2} \right)^2 }
\end{eqnarray}
where $\Omega$ is the full width at half maximum (FWHM) of the angular distribution of crystallites (called \textit{mosaicity} or \textit{mosaic spread}).

The reflectivity is derived  from equation (\ref{eq:I_h_mos}), with its peak value obtained for $\Delta \theta = 0$:
\begin{eqnarray}
R^{\text{peak}}_{mos} = {I^{\text{peak}}_{h,mos}(T_0) \over I_0(0)} = {1\over 2}{\left( 1 - e^{ -2W(0) Q T_0} \right) \, e^{-\mu T_0 \over \cos \theta}}
\label{eq:peakR_mos}
\end{eqnarray}
The thickness maximizing the peak reflectivity is derived from (\ref{eq:peakR_mos}):
\begin{eqnarray}
{\partial R^{\text{peak}}_{mos} \over \partial T} = 0\; \Leftrightarrow T_0 = {\ln \left( {2 W(0) Q \over \mu} +1 \right) \over 2 W(0) Q }.
\label{eq:EpOpt_mos}
\end{eqnarray}
Halloin \& Bastie [2005] propose a comprehensive version of the calculation of the diffracted intensity in a mosaic crystal. Specifically the equations to compute the structure factor (including atomic form factor and Debye-Waller factor) and a numerical method to calculate the $B_0$ function.

\subsection{Crystals having curved diffracting planes}

CDP crystals can be obtained in three ways [Smither et al., 2005]. One is by applying a thermal gradient perpendicular to the considered planes of a perfect single crystal. The second is by bending elastically a perfect single crystal. This bending can be obtained thanks to an external device applying a force on the crystal (as it is commonly made for monochromators in synchrotron radiation facilities), but also through the deposition of a coating on a wafer or by grinding or grooving one face of a wafer.The third way is by growing a two-component crystal whose composition varies along the crystal growth axis.

Thermal gradient gives excellent results because it produces a very pure spherical curvature, but it requires a significant amount of power which is not available onboard a space borne observatory. Elastic bending by surface treatment seems promising for a Laue lens but it has not yet been investigated. Our CDP crystals were obtained by growing composition gradient crystals, which produces intrinsically curved planes without any mechanical stress. 

Equations describing the diffraction in such crystals are given by Malgrange [2002]. They are an extension of the PPK theory of diffraction in distorted crystals [Penning \& Polder, 1961; Kato, 1963] for the case of a large and homogeneous curvature. In this theory the distortion of diffracting planes is described by the strain gradient $\beta$:
\begin{eqnarray}
\beta =  {  { \Lambda_0 \over \cos^2 \, \theta_B } \: { \partial^{2} \vec{h} . \vec{u} \over \partial s_0 \, \partial s_h }   }.
\label{eq:def_beta0}
\end{eqnarray}
which can be written in a simpler way in the case of a uniform curvature:
\begin{eqnarray}
\beta = {\Omega \over T_0 \delta_w}
\end{eqnarray}
$\Omega$ being the FWHM of the angular distribution of planes, or mosaicity$^{\ref{fn:mosaicity}}$, $T_0$ the thickness of the crystal and $\delta_w$ half the Darwin width (Darwin width is defined as $2 \delta_w$). 

When the strain gradient becomes larger than a critical value $\beta_c = {\pi \over 2 \Lambda_0}$, Balibar et al. [1983] showed that a new wavefield is created which decreases the intensity in the diffracted wavefield. In the case of a uniform curvature of planes, when the condition $\beta > \beta_c$ is fulfilled the intensity diffracted at the plateau is given by Malgrange's formula [Keitel et al., 1999; Malgrange, 2002]:
\begin{eqnarray}
I_{h,curved} &=& I_0 \, \left( 1 - e^{-2 \pi \,{ \beta_c \over |\beta|}} \right) \, e^{-\mu T_0 \over \cos \theta_B}  \\
 &=& I_0 \, \left( 1 - e^{-\pi^2 \over \alpha} \right) \, e^{-\mu T_0 \over \cos \theta_B}  \label{eq:Icourbe}
\end{eqnarray}
The $\alpha$ parameter appearing in (\ref{eq:Icourbe}) can be expanded as 
\begin{eqnarray}
\alpha = {\pi \over 2}{|\beta| \over \beta_c}  = {\partial \theta / \partial T \, \Lambda_0 \over \delta_w} = {\Omega \,\Lambda_0^2 \over T_0 \, d_{hkl} }.
\label{eq:alpha_curved}
\end{eqnarray}
It is interpreted as the angular variation of the diffracting planes orientation over the extinction length, in unit of Darwin width.

The thickness maximizing the reflectivity can easily be obtained under the hypothesis that the curvature of planes $c_p$ is uniform. In this case  $c_p$ can be related to the mosaicity and the thickness as follows:
\begin{eqnarray}
c_p = {\Omega \over T_0}
\label{eq:def_curvature}
\end{eqnarray}
Inserting (\ref{eq:def_curvature}) in (\ref{eq:Icourbe}), one gets
\begin{eqnarray}
I_{h,curved} = I_0 \, \left( 1 - e^{-{\pi^2 d_{hkl} \over c_p \, \Lambda_0^2}} \right) \, e^{-{\mu \, \Omega \over c_p \, \cos \theta_B}}.
\end{eqnarray}
Let us define 
\begin{eqnarray}
M = {\pi^2 d_{hkl} \over \Lambda_0^2}  \quad \text{and} \quad N= {\mu \, \Omega \over \cos \theta_B}.
\end{eqnarray}
It follows that the curvature of planes maximizing the reflectivity is obtained by solving the equation:
\begin{eqnarray}
{\partial I_h \over \partial c_p} = 0 \;  &\Leftrightarrow& \; {\partial \over \partial c_p} \left[ \left( 1 - e^{- {M \over c_p}} \right) \, e^{-{N \over c_p}} \right]  = 0 \\
 &\Leftrightarrow& \; c_p^{opt} = {M \over \ln \left( 1 + {M \over N} \right) }.
\end{eqnarray}
Using (\ref{eq:def_curvature}) we obtain the thickness maximizing the reflectivity as a function of energy and mosaicity (for a given crystal and reflection):
\begin{eqnarray}
T_0 = {\Omega \,  \ln \left( 1 + {M \over N} \right) \over M}.
\label{eq:Topt_curve}
\end{eqnarray}

\subsection{Comparison}

There are two main differences between diffraction properties of mosaic and CDP crystals.
Firstly, the diffraction efficiency is limited to 50\% in the case of mosaic crystals while it can reach 100\% with CDP crystals. Secondly the diffraction profile of a mosaic crystal is close to the Gaussian while it is rectangular for a CDP crystal. The spatial distribution of the beam diffracted by a crystal being the convolution of its diffraction profile by its spatial extent (assuming a uniform beam larger than the crystal), the footprint of a CDP crystal onto the focal plane does not have the large tails of a Gaussian. In other words CDP crystals better concentrate signal than mosaic crystals. These two points together make a sizable difference in the resulting sensitivity of a telescope. In fact, a telescope having its lens composed of CDP crystals would be 75 \% more sensitive than the same telescope which lens would be composed of mosaic crystals (considering identical materials) [Barri{\`e}re, 2008].

\begin{figure}
\begin{center}
\includegraphics[width=1.04\textwidth]{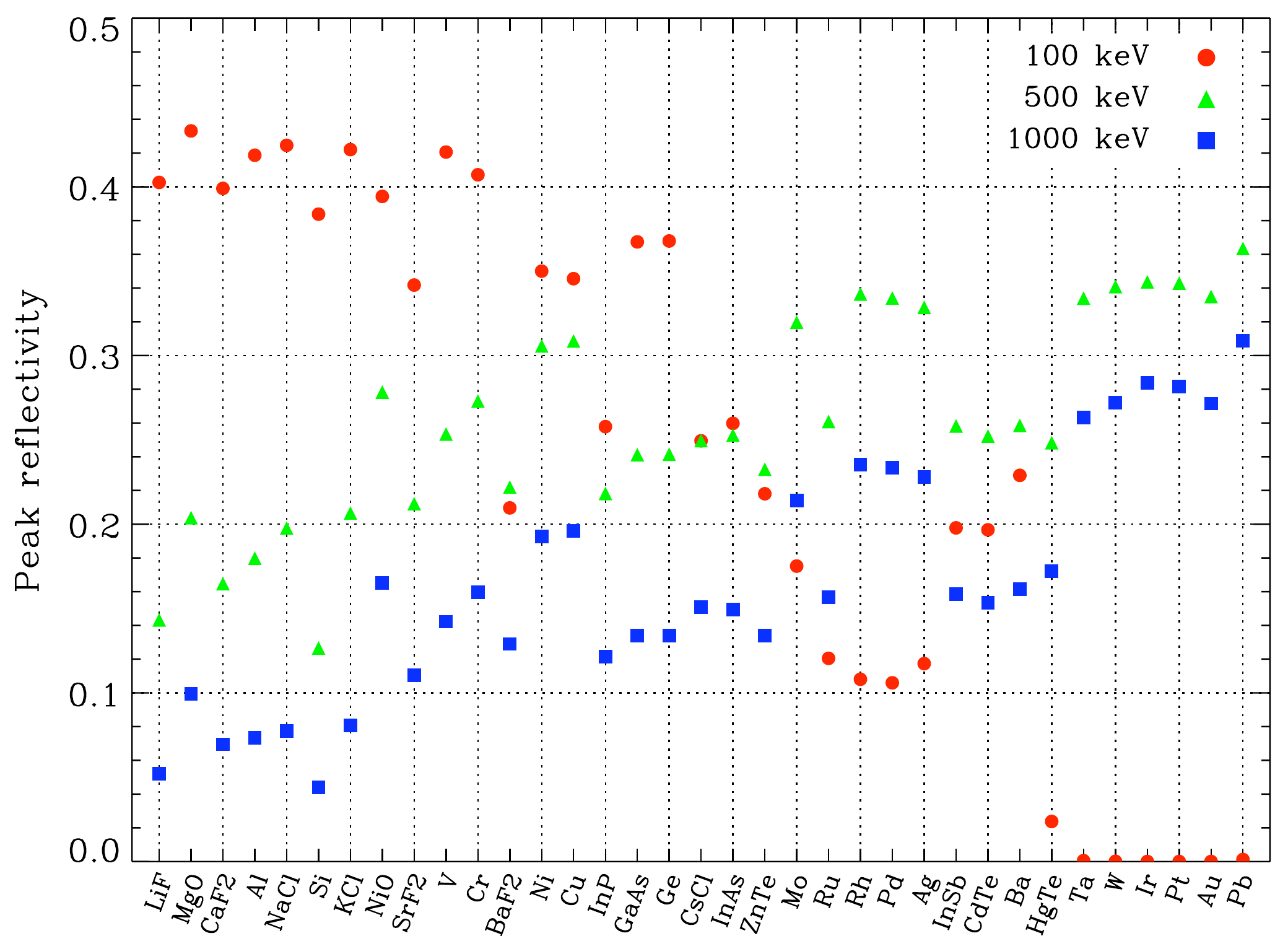}
\caption{Peak reflectivity of various potentially intresting crystals, sorted by increasing mean atomic number (Z) (and density in case of equal mean Z). The calcul assumes mosaic crystals with a mosaicity of 30 arcsec, a mean crystallites size of 5 $\mu$m and a thickness optimized to maximize the peak reflectivity within the limits 1 mm $\leqslant$ $T_0$ $\leqslant$ 25 mm. For each crystal, the reflection considered is given in Table \ref{tab:InfoCryst}.}
\label{fig:Reflectivity}
\end{center}
\end{figure}

Hence CDP crystals are substantially more suitable than mosaic crystals for the realization of a Laue lens, but in reality they are much more difficult to produce. Si$_{1-x}$Ge$_x$ are the only ones that we have been able to procure up until now. A feasibility study concerning Ge$_{1-x}$Sn$_x$ has been carried out at the Institute of Crystal Growth (IKZ, Berlin, Germany) which concluded that it was not feasible. Indeed the maximal solubility of Sn in Ge is about 1\% (instead of 100\% for Ge in Si) which strongly limits the range of the composition gradient. Moreover, the growth of such crystal is extremely difficult because of constitutional under-cooling on the liquid-solid interface, so that the 1\% maximum concentration cannot be reached experimentally.
Another option that has not been explored yet is Ni$_{1-x}$Sn$_x$ (mentioned in [Smither et al., 1982]).  Seen the reflectivity of Ni (Figure \ref{fig:Reflectivity} in the next section) the potential of such a composition gradient crystal appears very appealing.

\section{Search for suitable crystals: theoretical study}     
     
Our field of investigation has been limited to pure materials and two-component crystals. Crystals composed of more than two elements have not been considered because their large crystal cell volume strongly decreases their diffracted intensities. We will start with the selection of potentially interesting pure materials.

\begin{figure}
\begin{center}
\includegraphics[width=0.95\textwidth]{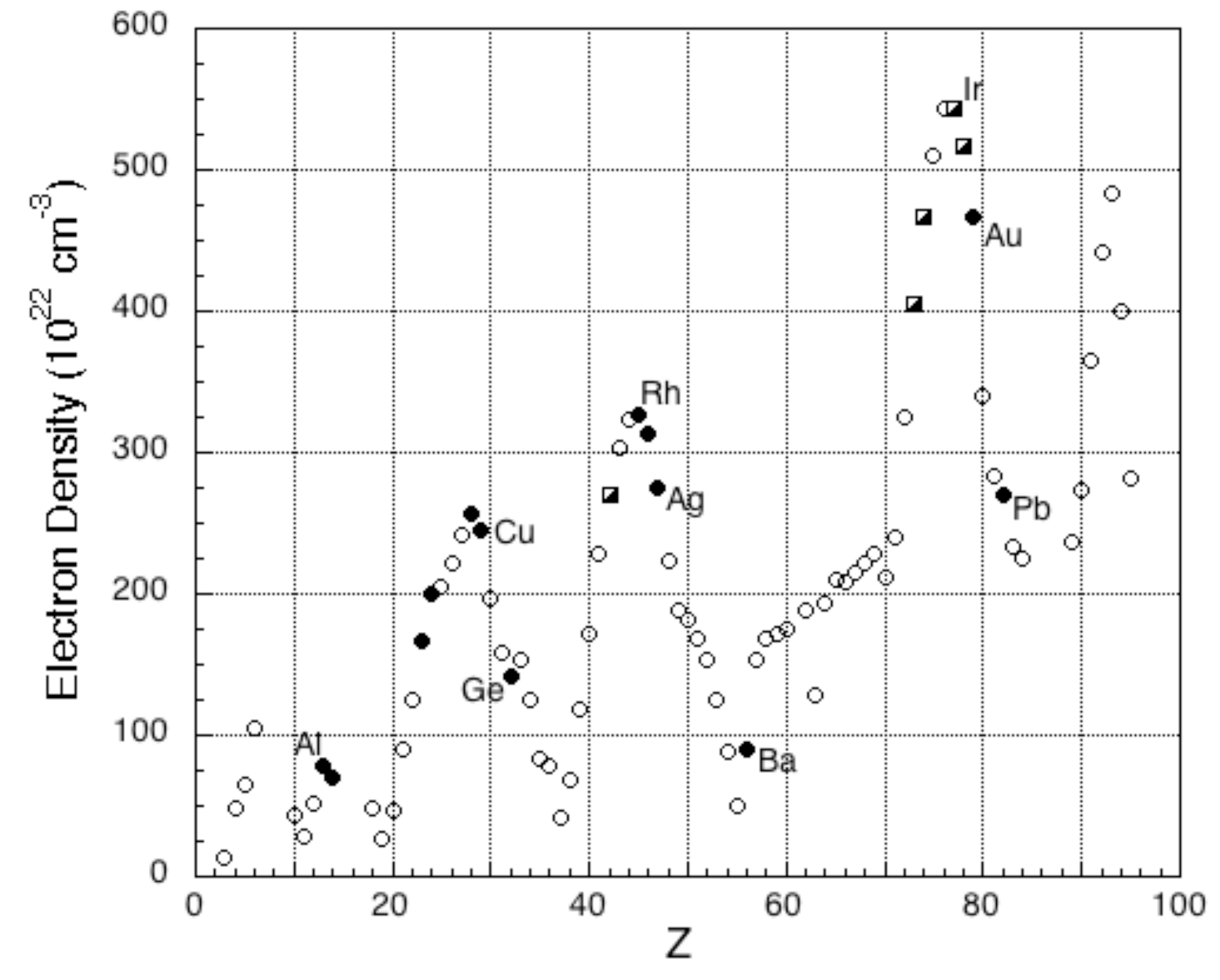}
\caption{Electron density in crystalline matter.}
\label{fig:Electrondensity}
\end{center}
\end{figure}

Suitable materials to realize a Laue lens must first and foremost exist at crystalline state at ambient temperature and pressure conditions without being too reactive in air (spontaneous combustion, deep oxidation) nor radioactive and overly toxic. Secondly they must diffract efficiently X-ray radiations, which is linked to a high electron density and depends also on the crystal lattice: the most efficient crystals have either a diamond (diam), a face-centered cubic (fcc) or a body-centered cubic (bcc) lattice. Figure \ref{fig:Electrondensity} shows the electron density of pure elements in crystalline state. Half-filled squares represent elements fulfilling the above conditions but having a melting point above 2000 $^{\circ}$C and/or being rare and expensive. Filled circles represent elements fulfilling the conditions and being more readily available than the ones represented as half-filled squares.

Eighteen out of about one hundred elements are potentially interesting, namely: Al, Si, V, Cr, Ni, Cu, Ge, Mo, Rh, Pd, Ag, Ba, Ta, W, Ir, Pt, Au and Pb. Among these eighteen elements numerous are soft and/or ductile as for instance Cu, Ag, Au, Pb. Crystals must be mechanically robust enough to undergo an accurate orientation and must not deteriorate during the intense vibrations induced by the rocket launch. Soft materials are nevertheless kept in the list because a slight doping is often enough to modify mechanical properties without changing the X-ray diffraction properties.

Another important selection criterion is the availability in large quantities with a good homogeneity. Crystals whose growth is well mastered are of prime interest. Unfortunately applications of pure-element crystal are rather limited, which explains why there are almost no industrial patterns to get a large quantity of constant-quality ingots, at the exception of Ge and Si. On the other hand two-component crystals are used in many applications like for instance GaAs, InAs, InP in electronics, CaF$_2$ for UV lithography, or CdTe for hard X-ray detectors, etc... Therefore we have added to the eighteen pure elements an arbitrary selection (still open to new entries) of two-component crystals easily available in industry that may represent an interesting perspective for the realization of a Laue lens. An extra advantage of two-component crystals is that some of them cleave. To have the external faces representative of crystalline planes can be a strong asset for the mounting and orientation of a large numbers of crystals on a lens structure.

In order to quantify and compare the diffraction capability of these crystals we have calculated their peak reflectivity for three different energies covering our domain of interest (100 keV, 500 keV and 1 MeV) assuming they are mosaic crystals. Darwin's model using the dynamical theory (\ref{eq:peakR_mos}) has been employed considering that the mosaicity is 30 arcsec, the crystallites thickness is 5 $\mu$m and the thickness is calculated to maximize the peak reflectivity according to equation (\ref{eq:EpOpt_mos}), within the limits 1 mm $\leqslant$ $T_0$ $\leqslant$ 25 mm. In each case the most intense reflection is considered as stated in Table \ref{tab:InfoCryst}.

Figure \ref{fig:Reflectivity} shows the results. As expected at high energy the crystals having the highest reflectivity are the ones with a high mean atomic number (Z) (see Table \ref{tab:InfoCryst}), and conversely at low energy the highest reflectivity is produced by crystals having a low mean Z. For the low energy side, this effect is due to the fact that the minimum thickness allowed in the calculation is 1 mm, which makes high mean Z crystals very absorbent at 100 keV. Various crystals both pure or two-component can yield a reflectivity above 40\% at 100 keV. In the scope of this study the best at 100 keV is MgO, which have two advantages: it cleaves along (100) planes, and it is very hard (hardness of 6 on Mohs scale). However the possibility to find it with a suitable mosaicity is still an open question.

Crystals with hexagonal compact (hc) lattice have not been regarded in this study because this lattice is not as efficient as the cubic one. As an example $_{44}$Ru (hc) and $_{45}$Rh (fcc) have almost an identical density of electrons (Figure \ref{fig:Electrondensity}) but Ru is far less efficient than Rh at 500 keV and 1 MeV. 

At 500 keV we see that the two-component crystals are not any more a good option. Oppositely Cu, Ni, Ag, Rh, and Pb exhibit high reflectivities. They are relatively affordable and produced with a mosaic structure (private communication), which makes them good potential candidates.  At 1 MeV, in the conditions taken for this calculation, only Ta, W, Ir, Pt, Au and Pb are yielding a reflectivity above 25\%, Pb being the most efficient bearing more than 30\%. The group Ta, W and Ir have very elevated melting points (2996 K, 3410 K and 2410 K respectively) causing  them to be expensive and difficult to get with a constant quality. Pt is very expensive and hence not produced in industrial quantities (several kilograms). At the moment remains only Pb and Au as viable candidates. Au is quite expensive but still affordable and seems to be available with a suitable mosaicity, as shown in the experimental results.

This study highlights that a careful choice of crystals can permit to cover from 100 keV to up to 1 MeV with a reflectivity above 30 \%, which is outstanding for this energy domain. It is necessary to identify many candidates for the same energy sub-range in order to then use only those whose actual measured quality is satisfying our requirements.

\section{Experimental results and discussion}
%
\subsection{Facilities and data treatment}

Since 2005 numerous samples have been measured using two facilities in Grenoble (France): the GAMS 4 instrument at ILL and the beamline ID15A at the European Synchrotron Radiation Facility (ESRF).

At ESRF we used two Ge 711 monochromators bent at the Rowland circle to get a sharp monochromaticity between about 280 keV and 600 keV with a fixed exit. Our samples were held by a sucking plate specially designed to allow fast sample changes without damage. The measurements consist in rocking curves (RCs) in Laue geometry both in diffraction and transmission. RCs are performed one after the other with a single high-purity Ge detector which is moved from one beam to the other. The beam intensity is monitored by the current of electrons in the storage ring of the synchrotron.

The GAMS 4 instrument uses the gamma-ray flux produced by neutron capture in a target inserted close to the nuclear reactor of ILL. We used an erbium target to produce lines among which was the one at 815.986 keV ($\Delta E / E \sim 10^{-6}$). To select the line we used a low mosaicity quartz crystal that gave a beam divergence of about 2 arcsec. The same type of detector and method as in ESRF was used to record the RCs.

RCs in transmission and diffraction recorded on the same area of the crystal samples are put together and normalized by the intensity of the transmitted beam when no diffraction occurs.  Thus complementary peak and valley show directly the diffraction efficiency. Reflectivity is then obtained by applying the transmission coefficient to take into account the absorption through the crystal.

In the case of mosaic crystals RCs are fitted using Darwin's model (as defined in section \ref{sec:MosCryst}) which allows the extraction of the mosaicity and of the crystallites size. To be coherent the diffraction efficiency values presented in this paper also come from the fit of Darwin's model. In the case of CDP crystals, an average of the diffraction efficiency is made over the width of the plateau.

\subsection{Copper mosaic crystals}

Copper crystals are produced by the monochromator group of ILL, where they manage the growth of 8 kg ingots of very low mosaicity [Courtois et al. 2005]. The group has extensive experience in growing high quality Cu crystals, but our requirements in mosaicity are far below their usual needs. Mosaicity and homogeneity were the main challenge with Cu mosaic crystals. Lately, some pieces of Cu mosaic crystal featuring a mosaicity below 1 arcmin have been measured. Three examples are shown hereafter (all samples have been measured as-cut without surface treatment).

\begin{figure}
\includegraphics[width=\textwidth]{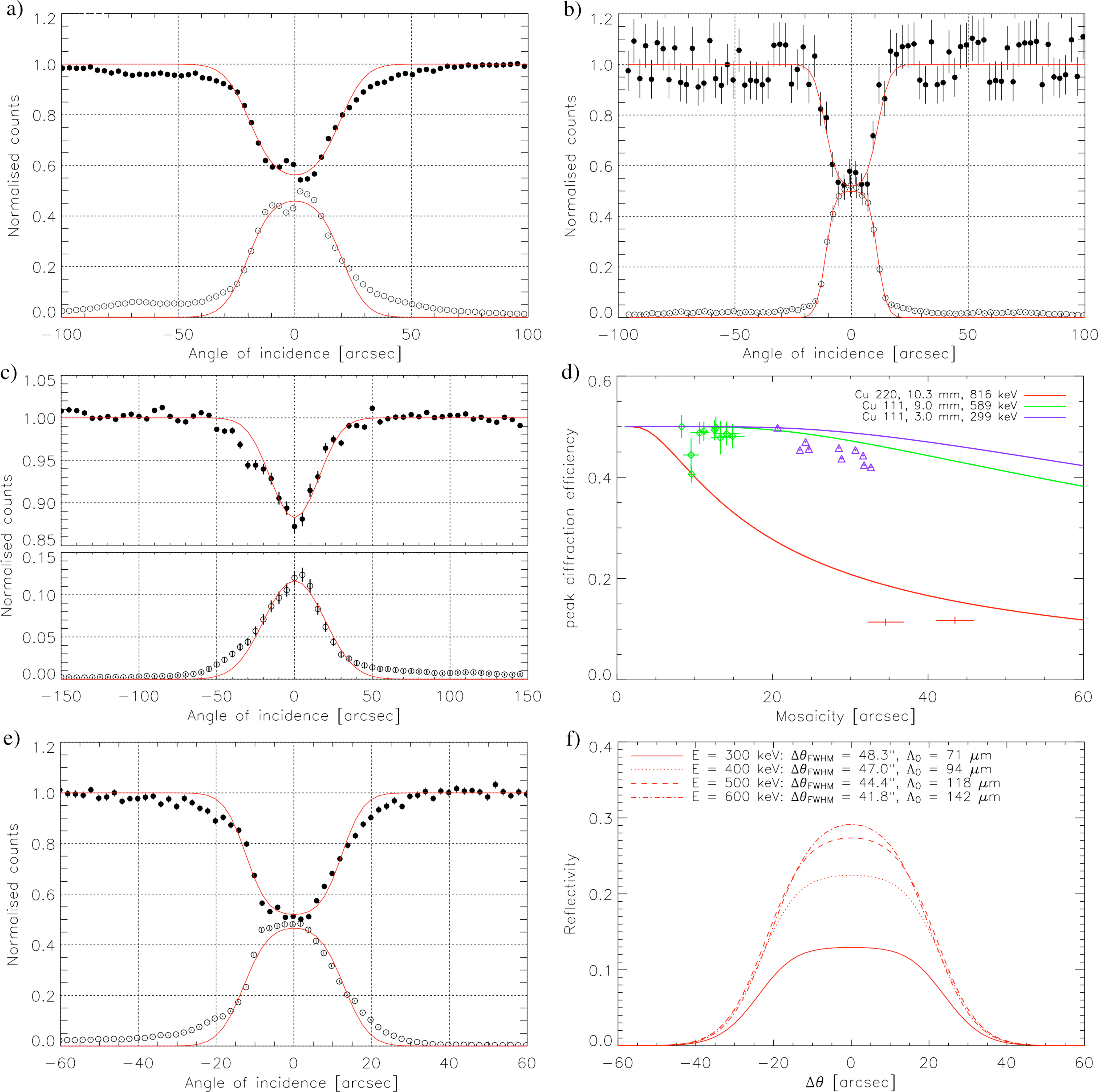}
\caption{\textbf{a)} RCs recorded on the crystal Cu 834.$\delta$4.3 using a beam of 299 keV and the 111 reflection. The crystal is 3 mm thick, surface has been left as cut which may be the reason of the tails around the peak. The mosaicity in this point is 29 arcsec and the diffraction efficiency reaches 47\% yielding a reflectivity of 35\% taking into account the transmission coefficient of 0.75.
\textbf{b)} RCs recorded on the crystal Cu834.$\delta$3.2 using a beam of 589 keV and the 111 reflection. Crystal thickness equals 9 mm, which makes a transmission coefficient of 0.55. Mosaicity in this point is 14 arcsec. The diffraction efficiency reaches 50\%, yielding a reflectivity of 27.5\%.
\textbf{c)} Averaged RCs over ten measurement spots on crystal Cu 834.31 using the 220 reflection and a 816 keV beam. 
\textbf{d)} Peak diffraction efficiencies measured on the three Cu crystals 834.31 (averaged over the ten spots), 834.$\delta3$ and 834.$\delta4$ as a function of mosaicity. Experimental results are compared to Darwin's model using the kinematical theory.
\textbf{e)} RCs recorded on 111 reflection of the gold sample Au\_1 with a beam energy of 588 keV. The estimated mosaicity is 16 arcsec. Taking into account the absorption through the thickness of 2 mm a reflectivity of 31\% is found. Continous line indicates a fit to Darwin's model.
\textbf{f)} Simulated rocking curves using Darwin's model of an Au crystal (111 reflection) of 2 mm thick, 30 arcsec of mosaicity and having crystallites of 40 $\mu$m of thickness.}
\label{fig:mos_cryst}
\end{figure}

The first one is the sample 834.$\delta4$ (55 x 20 mm$^2$ cross section and 3 mm thick). It was measured at ESRF in 25 points using the 111 reflection in a 299 keV beam. Averaged values are reported in the Table \ref{tab:Cu} and examples of RCs  are shown in Figure \ref{fig:mos_cryst} a). As we can see Darwin's model (continuous line) fits quite well the data. Note however that the RCs show small enlargements at the base most likely due to defect induced by the cut. This sample shows a good homogeneity with an average mosaicity of 25 arcsec and a standard deviation of 6 arcsec only (over 25 spots). Its average diffraction efficiency is 46 \%, which corresponds to an excellent reflectivity of 34 \%. This sample is an excellent example of what is needed to build a Laue lens.

The second sample, 834.$\delta3$ (55 x 20 mm$^2$ cross section and 9 mm thick), has been measured at ESRF in 25 spots using the 111 reflection in a 589 keV beam. The average parameters extracted from the data are summarized in the Table \ref{tab:Cu} and examples of the RCs are shown in Figure \ref{fig:mos_cryst} b). Despite instrumental problems that produced an oscillation of the counting in transmission geometry we immediately notice that Darwin's model fits very well and that these RCs are very narrow for a Cu crystal. Indeed the average mosaicity of this sample is 14 $\pm$ 6 arcsec which is among the smallest values ever measured in a Cu crystal. Initially the thickness was optimized for 850 keV which explains that despite a good diffraction efficiency the reflectivity is 'only' 26\%.

The last Cu sample, Cu 834.31 (19.2 x 21 mm$^2$, 10.3 mm thick), has been measured in the 220 reflection at 816 keV (GAMS 4) over ten large spots of 2 mm x 11 mm covering more than half its face. RCs shown in Figure \ref{fig:mos_cryst} c) are averaged over all spots. As indicated in Table \ref{tab:Cu} the mean mosaicity is 39 arcsec and its standard deviation over the ten spots is 6 arcsec which indicates a satisfactory homogeneity. 

Disregarding the influence of the crystallites thickness (i.e. in the limit of the kinematical theory) the diffraction efficiency decreases with $W(0)Q_{kin}$ which is proportional  to ${d_{hkl} |F_{hkl}|^2 \,/ \, E^2 \, \Omega}$. Knowing that $d_{220} < d_{111}$ and $|F_{220}|^2 < |F_{111}|^2$ one understands why this crystal yields a lower diffraction efficiency than the two former, equaling 12\%.
Figure \ref{fig:mos_cryst} d) shows the measured peak diffraction efficiency versus the mosaicity for the three Cu samples and for comparison theoretical values according to the kinematical theory (calculated using eq. (\ref{eq:peakR_mos}) with a crystallites thickness of 0.1 $\mu$m but disregarding the absorption term). As can be seen, at 816 keV even an ideally imperfect crystal is not efficient with the 220 reflection: with a mosaicity of 40 arcsec, we could have expected at most a diffraction efficiency of 16.3\%(which would have resulted in a reflectivity of 9.1\%). For comparison, in the same conditions the 111 reflection would lead to a diffraction efficiency of 35.7\%, making a reflectivity of 19.7\%.

It results from this study that Cu crystals as produced at ILL are usable for the realization of a Laue lens. Between energies of $\sim$ 300 keV and $\sim$ 600 keV these crystals give optimal results. However during our experimental runs we have noticed that cutting is critical to the final properties of the crystal sample (performed with an EDM machine at ILL). It can induce defects over depths of the order of 1 mm or more which cannot be removed by acid etching. A systematic study would have to be undertaken to determine the best compromise between cutting speed and degradation of the crystals.

\subsection{Si$_{1-x}$Ge$_x$ concentration gradient crystals}

Si$_{1-x}$Ge$_x$ alloy with $x$ increasing along the crystal growth axis (gradient crystal) is produced at IKZ [Abrosimov, 2005].  The increase of Ge concentration deforms the silicon lattice resulting in a spherical curvature of the diffracting planes perpendicular to the growth axis. In our case, crystals have been grown along the [111] direction, which results in a spherical curvature of the (111) planes. Since the Ge concentration remains low in the crystal the curvature is approximatively proportional to the gradient of Ge concentration [Smither et al., 2005]:
\begin{eqnarray}
\nabla C_{Ge} \approx \epsilon \,/\, R_c
\end{eqnarray}
where $\nabla C_{Ge}$ is the gradient of Ge concentration expressed in atomic percentage per centimeter, $R_c$ the radius of curvature of planes (in meters), and $\epsilon$ a constant. The assumption that the curvature is regular (spherical) permits to relate the gradient of Ge concentration to the mosaicity $\Omega$ 
\begin{eqnarray}
\nabla C_{Ge} \approx \epsilon \,\Omega / T_0
\label{eq:CGe_Omega}
\end{eqnarray}
with $T_0$ the thickness of the crystal. $\epsilon$ equals 25 in Smither's paper.

\begin{figure}
\includegraphics[width=\textwidth]{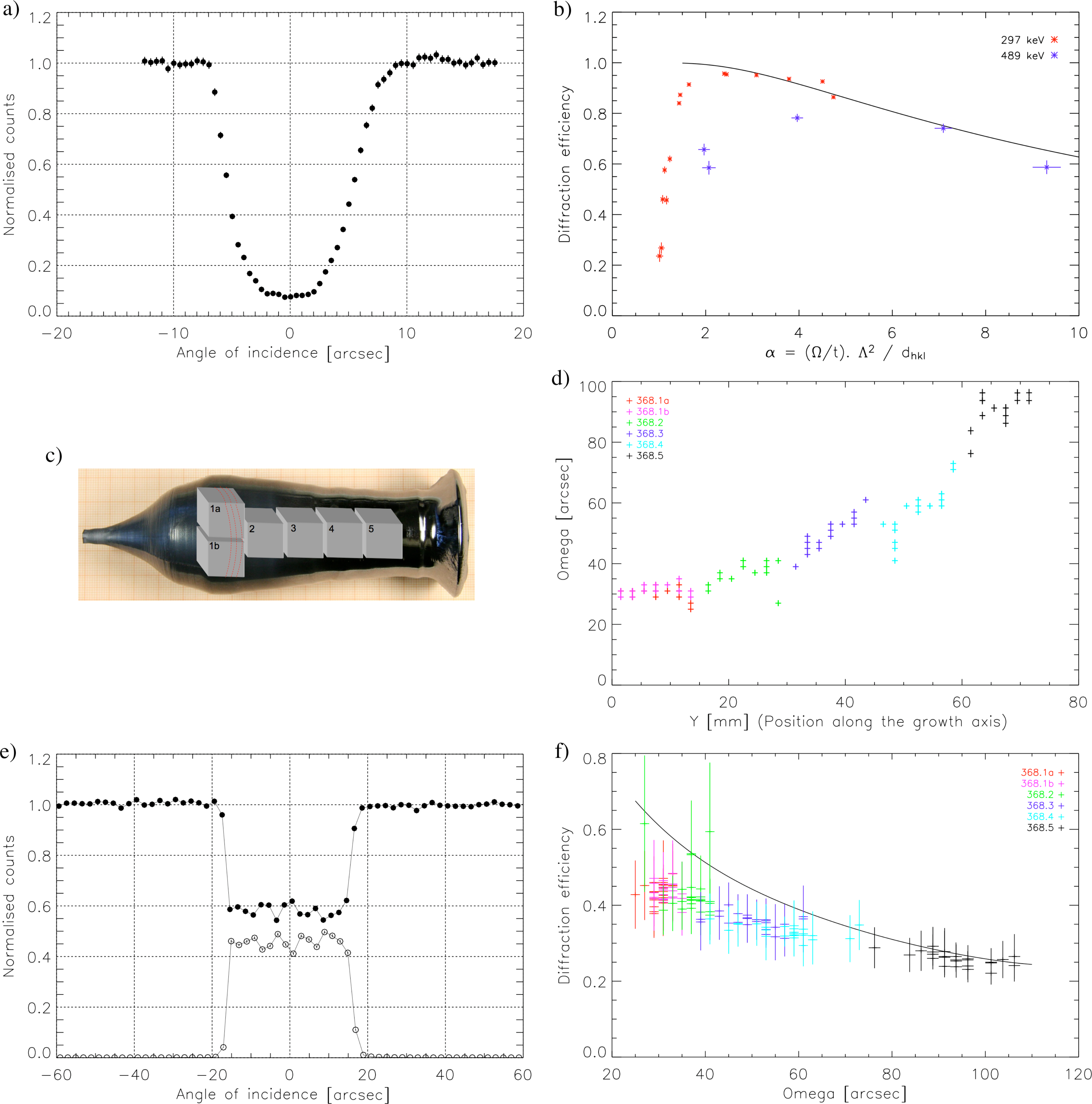}
\caption{
\textbf{a)} One of the most striking RC recorded on the sample SiGe 10.3 from 111 reflection. The thickness of the sample is 20 mm, and the beam energy is 297 keV. The mosaicity equals 12 arcsec and the reflectivity 60\%. 
\textbf{b)} Diffraction efficiency measured on crystal SiGe 10.3 at 297 keV (red) and at 489 keV (blue), along the growth axis. The continuous line shows the theoretical prediction assuming a constant Ge concentration of 2.5 \%. 
\textbf{c)} Ingot 368 of Si$_{1-x}$Ge$_x$, x varying along the growth axis (x increase towards the right in the picture), produced at IKZ. Six pieces each measuring 15 x 15 x 23 mm$^3$ were extracted (as shown in the figure). The dashed line represents the  (111) spherical planes. 
\textbf{d)} Mosaicity as a function of crystal position along the growth axis in the six pieces extracted from ingot SiGe 368. Measurements have been performed at ESRF using the 111 reflection and a 299 keV beam. 
\textbf{e)} A very good example of a measured RC from the 111 reflection of sample SiGe 368 1b: the rectangular shape of the curves is very close to what is expected in the ideal case, showing the regularity of the curvature of diffracting planes. The thickness of the sample is 23 mm and the beam energy was 299 keV. 
\textbf{f)} Peak diffraction efficiency measured on the six pieces extracted from ingot SiGe 368 as a function of mosaicity and compared to theoretical predictions (continuous line). 
}
\label{fig:bent_cryst}
\end{figure}

The first part of the investigation consisted in the characterization of the diffraction properties of several spare samples given by IKZ of which SiGe 10.3 is of particular interest. This sample (28 mm x 16 mm, 20 mm thick) has been measured twice at ESRF using the 111 reflection. The first time we measured 15 points on the crystal axis (along the gradient direction) using a 297 keV beam. We noticed that the mosaicity increased from the low Ge concentration side to the high Ge concentration side as expected. The change in curvature is a convenient tool to study the accordance with the theory. Among the RCs recorded the most outstanding is shown in Figure \ref{fig:bent_cryst} a).

This sample has also been measured in 5 points along the growth axis using a 489 keV beam. Results of both series of measurements have been gathered in Figure \ref{fig:bent_cryst} b) where the peak (or plateau) diffraction efficiency is plotted as a function of $\alpha$ (\ref{eq:alpha_curved}), a variable that increases both with $E$ and $\Omega$. The black continuous line shows the theoretical predictions according to (\ref{eq:Icourbe}) using as input a mean Ge concentration of 2.5 at.\% assumed constant over the 16 mm width of the sample (it intervenes in the calculation of the d-spacing, the structure factor and the absorption factor). Accordance with theory seem to appear for larger $\alpha$ values when energy increases. This point has been noticed on other sample as well, but is not yet fully understood.

In 2007 three ingots of SiGe were produced in IKZ with the objective of checking the relation between growth parameters (initial Ge concentration, pulling speed, shape of the crystal) and gradient of Ge concentration, which is evaluated through the RC width thanks to equation (\ref{eq:CGe_Omega}). In this series the ingot referred to as SiGe 368 has allowed the extraction of six pieces of 15 x 15 x 23 mm$^3$ as represented in Figure \ref{fig:bent_cryst} c). The pieces have been used as-cut without further surface treatment as it has been previously established that nothing changed after a deep acid etching (0.5 mm on each face). Every piece was measured in 13 different spots at ESRF using a 299 keV monochromatic beam and the 111 reflection (RC in transmission  and diffraction as shown in Figure \ref{fig:bent_cryst} e)). Figure \ref{fig:bent_cryst} d) shows the  RC width as a function of the position along the growth axis, the low Y side having the lowest Ge concentration. $\Omega$ increases from 30 arcsec up to about 95 arcsec over the 75 mm length of the various pieces, but is relatively constant in the two asymmetric pieces 1a and 1b.

Figure \ref{fig:bent_cryst} f) is a plot of the diffraction efficiency measured on each spot as a function of the RC width $\Omega$. On this plot the continuous line shows the theoretical prediction calculated from (\ref{eq:Icourbe}), but this time using an evolving Ge concentration: using the measured Ge concentration at the extremity of the samples (the low concentration side of samples 1a and 1b and the high concentration side of sample 5), and knowing the corresponding solidified fraction $g$ (see Table \ref{tab:SiGe368}) we refined the value of the segregation coefficient to $k$ = 0.42 (assuming it to be constant on this part of the crystal) and plugged it into the Scheil-Pfann formula:
\begin{eqnarray}
C_{Ge}^s= k \,  C_{Ge, \,0} \, (1-g)^{k-1}
\end{eqnarray}
where $C_{Ge}^s$ is the Ge concentration in the solid phase, and $C_{Ge, \,0}$ is the initial Ge concentration in the melt. $C_{Ge, \,0}$ = 0.05 in the case of ingot SiGe 368. With this data we can  approximate the Ge concentration quite well along the 75 mm of the 6 samples (see right column of Table \ref{tab:SiGe368}) and so determine the theoretical diffraction efficiency. 
The agreement with theory is quite good especially for the large values of $\Omega$, where the curvature of planes is larger. We could have plotted the diffraction efficiency versus $\alpha$ which would have spanned from 9 to 35 in this case. This would show, again, that the accordance is better for larger $\alpha$ values than for small ones. Unfortunately it has not been possible to measure the SiGe 368 samples at a different energy (we have already recorded 156 RC!). It would have permit to check if the 'accordance point' shifts towards larger $\alpha$ for a higher energy. 
The agreement between theory and experiments means that the curvature of diffracting planes is very regular. We also notice that there is no visible difference between symmetrical (2,3,4 and 5) and asymmetrical (1a and 1b) cut elements which bodes well for our application since it means that the whole volume of the ingot can be utilized.

The ultimate objective of this development process is to grow crystals having a constant gradient of Ge concentration which value would produce a curvature optimized for a RC width of 30 arcsec at 300 keV. 
All indications point towards this goal being achievable. New samples that have been produced in early Fall 2008 will hopefully confirm this during a beamtime slot at ESRF foreseen in Spring 2009.

\subsection{Prospective for high-reflectivity crystals: Au}

In the context of the search for more efficient crystals for high-energy diffraction, a gold crystal (a disc of 10 mm in diameter and 2 mm thick) has been purchased from the German company Mateck. This crystal, which has been polished, was measured at four different energies (299, 399, 494 and 588 keV) at ESRF during Spring 2008. For each energy five spots (forming a square of 5 mm of diagonal plus its center) of 1mm x 1mm were probed using the 111 reflection. Results have been compounded in Table \ref{tab:Au_1} and a typical RC of this crystal is shown in Figure \ref{fig:mos_cryst} e). 

Results are very positive; the crystal seems homogenous with a mosaicity ranging from 16 arcsec at 588 keV to 30 arcsec at 300 keV and achieve an excellent diffraction efficiency. This high-Z material was the first of a new wave of investigations. A less successful attempt was also conducted with a Pt crystal (from Mateck) which showed multiple grains and an overall mosaicity larger than 1 degree. Recently several Ag and Rh crystals have been purchased (from Mateck) and will be characterized soon.

\subsection{Limitation of Darwin's model}

In the previous example  (Table \ref{tab:Au_1}) the crystallites thickness seems to increase as the inferred mosaicity decreases. This apparently non-physical behaviour has been observed in other measurements of mosaic crystals [Barri\`ere, 2008] as well and is not explained by DarwinÕs model.
As shown in Figure \ref{fig:mos_cryst} f), it is expected that the FWHM of the RCs decreases when energy increases. It is due to the fact that the extinction length (\ref{eq:extlength}) is proportional to the energy: at low energy the extinction length is smaller than crystallites, which implies that the diffraction peak is cut due to primary extinction in each crystallite. The peak being cut, it results in a larger FWHM. Hence the fit of data with Darwin's model does not require to decrease the mosaicity to fit this FWHM falloff. But the observed effect is much more important than what can be modeled by dynamical effects in a crystal described by Darwin's model.

Moreover the apparent increase of crystallites size is hardly explained in the same way: if the apparent size of crystallites was driven by the extinction length the value of A (defined in eq. \ref{eq:defA}) should remain constant with respect to the energy of measurement, which is not the case. Our interpretation of these phenomena is that the crystallites in a mosaic crystal are not all identical but have a size distribution. Hence one emphasizes either the small crystallites when the beam energy is low, because large crystallites have a lower contribution due to primary extinction. On the other hand one emphasizes the large crystallites at high energy when the longer extinction length bring the crystallites in a regime where they diffract proportionally to their size (i.e. a regime described by the kinematical theory). In this idea, the observed decrease of mosaicity can be explained if one assumes that the smaller crystallites have a correspondingly larger angular distribution than the larger crystallites, which seems a reasonable hypothesis.

A consequence of the increase of crystallites size when the energy of measurement increases is that diffraction keep in the regime of dynamical theory instead of entering the regime of kinematical theory as we could have expected seen the increase of extinction length. It means that crystals are less efficient that what we could have expected from their parameters deduced from low-energy measurements. Accurate performance estimates of Laue lenses require a reliable modeling of intensities diffracted by crystals. Since the energy dependance of the crystal parameters is not yet modeled, it is not possible to extrapolate accurately the performance of a crystal from one energy to another. There are two possibilities to solve this problem, either we try to refine Darwin's model integrating a dispersion of crystallites size or we can try to determine the 'evolution' of the crystallites size that we have to input in Darwin's model to stick with experimental results. In the present dataset the evolution of $A$ with energy is very well fitted by a second degree polynomial curve (R$^2$ = 0.9994) with the parameters:
 \begin{eqnarray}
 A = 3.759 - 7.650\text{x}10^{-3} \,E + 6.382\text{x}10^{-6} \, E^2.
 \end{eqnarray}
But with only one crystal measured on one spot at four energies it is hard to derive conclusions. Is this evolution of $A$ the same for any crystal? Is this equation still valid if we extend the energy range of investigation? Either to refine Darwin's model or to model the evolution of $A$, in both cases more measurements are required.

\section{Conclusions - Perspectives}

The last iteration of Si$_{1-x}$Ge$_x$ production have given very satisfactory results, paving the road towards the final phase of our development program: the production of constant-gradient ingots allowing the extraction of  homogeneous pieces having a 30 arcsec bandpass and optimized for 300 keV. Cu crystals have also shown that our requirements are attainable with pieces having a mosaicity as low as 15 arcsec, and reflectivity in accordance with theoretical predictions.

A new phase in these investigations started with the characterization of a gold sample. This 2 mm thick sample has kept its promises with an excellent reflectivity near 600 keV. A Laue lens such as the one designed for the GRI mission could benefit dramatically from the enablement of high-Z crystals such as gold. In the coming months Ag and Rh crystal samples will be measured as well, hopefully enlarging our portfolio of usable materials, which is of prime importance for the design of efficient lenses.

More generally the identification of efficient crystals opens the way for additional experimental tests. Two-component crystals in particular are of great interest since their growth has already been developed for other applications. High-Z pure elements are quite rare and expensive, but are mandatory when building an efficient lens covering energies higher than 600 keV. Alternatively, crystals having intrinsically curved diffraction planes can enhance dramatically the overall performance of Laue lenses. This field, which is largely unexplored so far, could also find some interesting application in high-energy monochromators (X-rays and neutrons) a fact that further strengthens the drive for initiating a development process.\\







\ack{Acknowledgements}
{The authors wish to thank the French Space Agency for its continued support, as well as the European Space Agency. NB and LN are grateful to ASI for the support of Laue lens studies through grant I/088/06/0. In addition, we give special thanks to our local contacts of beamline ID15A of ESRF, Thomas Buslaps, Veijo Honkimaki and John Daniels.}




\begin{table}
\begin{tabular}{lccccc}      
\hline
Crystal & Mean & Lattice & Most intense & Debye & Cleavage \\
 & Z & & reflection & temp. (K) & plane \\
 \hline
 LiF		& 6 		& NaCl 		& 200 	& 723 \footnotemark[1]     & (100)  \\
 MgO 	& 10 	& NaCl		& 200 	& 743 \footnotemark[2]     & (100)  \\
 CaF$_2$	& 12.66 	& CaF$_2$ 	& 220 	& 354 \footnotemark[3]     & (111)  \\
 Al		& 13		& fcc 		& 111 	& 428 \footnotemark[4]     &            \\
 NaCl	& 14 	& NaCl		& 200	& 290 \footnotemark[5]     & (100)    \\
 Si		& 14 	& diamant		& 220 	& 645 \footnotemark[4]     &               \\
 KCl	 	& 18 	& NaCl		& 200	& 235 \footnotemark[5]     & (100)    \\
 NiO		& 18 	& NaCl		& 200 	& 317 \footnotemark[6]     &               \\
 SrF$_2$	& 18.66	& CaF2 		& 220 	& 262 \footnotemark[3]     & (111)    \\
 V 		& 23 	& bcc 		& 110 	& 380 \footnotemark[4]     &               \\
 Cr 		& 24 	& bcc		& 110 	& 630 \footnotemark[4]     &                 \\
 BaF$_2$ & 24.66	& CaF$_2$ 	& 220 	& 196 \footnotemark[3]     & (111)    \\
 Ni 		& 28		& fcc			& 111 	& 450 \footnotemark[4]     &                \\
 Cu 		& 29 	& fcc			& 111 	& 343 \footnotemark[4]     &                \\
 InP 		& 32 	& ZnS		& 220 	& 500 \footnotemark[7]     &               \\
 GaAs 	& 32 	& ZnS		& 220 	& 264 \footnotemark[8]     &               \\
 Ge 		& 32 	& diamant		& 220	& 374 \footnotemark[4]     &                \\
 CsCl 	& 36		& bcc 		& 110	& 162 \footnotemark[5]     &                \\
 InAs 	& 41		& ZnS		& 220 	& 350 \footnotemark[9]     &               \\
 ZnTe	& 41		& ZnS		& 220 	& 198 \footnotemark[10]   &              \\
 Mo 		& 42		& bcc		& 110 	& 450 \footnotemark[4]     &                \\
 Ru 		& 44		& hc			& 002	& 600 \footnotemark[4]     &                \\
 Rh 		& 45		& fcc			& 111 	& 480 \footnotemark[4]     &                \\
 Pd 		& 46		& fcc			& 111	& 274 \footnotemark[4]     &                \\
 Ag 		& 47		& fcc			& 111	& 225 \footnotemark[4]     &                \\
 InSb 	& 50		& ZnS		& 220 	& 280 \footnotemark[11]   &  (100)   \\
 CdTe 	& 50		& ZnS		& 220 	& 145 \footnotemark[12]   &               \\
 Ba 		& 56		& bcc		& 110 	& 110 \footnotemark[4]      &                \\
 HgTe	& 66		& ZnS		& 220 	& 105 \footnotemark[13]    &               \\
 Ta 		& 73	 	& bcc		& 110 	& 240 \footnotemark[4]       &                \\
 W 		& 74 	& bcc		& 110	& 400 \footnotemark[4]       &                \\
 Ir 		& 77 	& fcc			& 111	& 420 \footnotemark[4]       &                \\
 Pt 		& 78		& fcc			& 111	& 240 \footnotemark[4]       &                \\
 Au 		& 79 	& fcc			& 111	& 165 \footnotemark[4]       &                \\
 Pb 		& 82		& fcc			& 111	& 105 \footnotemark[4]       &                \\ 
\hline
\end{tabular}

\caption{Miscellaneous information about the crystals considered in this study. }
\label{tab:InfoCryst}
\end{table}

\footnotetext[1]{[Cotts \& Anderson, 1981]}
\footnotetext[2]{[Begg, 1976]}
\footnotetext[3]{[Palchoudhuri \& Bichile, 1989]}
\footnotetext[4]{[Kittel, 1970]}
\footnotetext[5]{[Kumara Swamy et al., 1996]}
\footnotetext[6]{[Freer, 1981]}
\footnotetext[7]{[Matsuo Kagaya \& Soma, 1986]}
\footnotetext[8]{[Stevenson, 1994]}
\footnotetext[9]{[Fawcett, 1969]}
\footnotetext[10]{[Bashir, 1988]}
\footnotetext[11]{[Racek, 1973]}
\footnotetext[12]{[Zub{\'\i}k, 1976]}
\footnotetext[13]{[Mavroides \& Kolesar, 1964]}

\begin{table}
\begin{tabular}{lccc}      
\hline
Crystal name & Cu 834.$\delta4$ & Cu 834.$\delta3$ & Cu 834.31 \\
Reflection & 111 & 111 & 220 \\
Thickness (mm) & 3.0 & 9.0 & 10.3 \\
Energy (keV) & 299 & 589 & 816 \\
Mosaicity (arcsec) & 25 $\pm$ 6 & 14 $\pm$ 6 & 39 $\pm$ 6  \\
Peak diffr. efficiency & 0.46 $\pm$ 0.02 & 0.47 $\pm$ 0.04 &  0.12 $\pm$ 0.02 \\
Peak reflectivity & 0.34 $\pm$ 0.02  & 0.26 $\pm$ 0.02 & 0.06 $\pm$  0.01 \\
Size of crystallites ($\mu m$) & 60 $\pm$ 15 & 129 $\pm$ 43 & 210 $\pm$ 31 \\
$A= \pi\, t_0 / \Lambda_0$ & 1.27 $\pm$ 0.32 & 1.38 $\pm$ 0.46 & 1.17 $\pm$ 0.17 \\
\hline
\end{tabular}
\caption{Summary of results of the Cu crystals investigated.
Mosaicity, diffraction efficiency and crystallites thickness are obtained from Darwin's model fits. Values are averaged over all measurement spots, and their uncertainty is the standard deviation. The bottom row allow the crystallites thickness to be related to the extinction length.}
\label{tab:Cu}
\end{table}

\begin{table}
\begin{tabular}{lccc}      
\hline
Y (mm) & g & $C_{Ge,\, exp}$ (at.\%) & $C_{Ge,\, calc}$ (at.\%) \\
\hline
 0   & 0.192 & 2.40 $\pm$ 0.15 &  2.38 \\
15 & 0.347 & - & 2.69 \\
30 & 0.484 & - & 3.08 \\
45 & 0.596 & - & 3.55 \\
60 & 0.692 & - & 4.16 \\
75 & 0.777 & 5.05 $\pm$ 0.20 & 5.01 \\
\hline
\end{tabular}

\caption{Position along the growth axis (Y), solidified fraction (g), measured Ge concentration ($C_{Ge,\, exp}$) and calculated Ge concentration ($C_{Ge,\, calc}$) in crystal SiGe 368.}

\label{tab:SiGe368}
\end{table}

\onecolumn
\begin{center}
\begin{table}
\begin{tabular}{lcccc}
\hline
Energy (keV) & 299 & 399 & 494 & 588 \\
Mosaicity (arcsec) &  30 $\pm$ 5 & 24 $\pm$ 3  & 24 $\pm$ 2 &  18 $\pm$3 \\
Peak diffraction efficiency & 0.47 $\pm$ 0.02  & 0.47 $\pm$ 0.01  & 0.46 $\pm$ 0.02 & 0.45 $\pm$ 0.02 \\
Peak reflectivity &  0.12 $\pm$ 0.01 &  0.22 $\pm$ 0.01 & 0.26 $\pm$ 0.01 & 0.29 $\pm$ 0.02 \\
Size of crystallites ($\mu m$) &  46 $\pm$ 3 & 52 $\pm$ 3 & 57 $\pm$ 4 & 65 $\pm$ 6\\
$A= \pi\, t_0 / \Lambda_0$ & 2.04  $\pm$ 0.13 & 1.73  $\pm$ 0.09 & 1.53  $\pm$ 0.11 & 1.47  $\pm$ 0.13 \\
\hline
\end{tabular}

\caption{Summary of the results obtained with gold crystal Au\_1 using the 111 reflection.
Mosaicity, diffraction efficiency and crystallites size are obtained from Darwin's model fits. Values are averaged over five measurement spots (1mm x 1mm), and their uncertainty is the standard deviation. Bottom row allows comparison of the crystallites size with the extinction length.}
\label{tab:Au_1}
\end{table}

\end{center}
\twocolumn


\onecolumn
\twocolumn

\onecolumn
\twocolumn

\end{document}